\documentclass[sigconf, screen]{acmart}

\graphicspath{{figures/}}

\AtBeginDocument{
  }

\setcopyright{none}

\begin{document}

\title[Contact-Free Sensing of Human HR and Canine BR for AAIs]{Contact-Free Simultaneous Sensing of Human Heart Rate and Canine Breathing Rate for Animal Assisted Interactions}

\settopmatter{printacmref=false}
\renewcommand\footnotetextcopyrightpermission[1]{}
\pagestyle{plain}

\author{Timothy	Holder}
\affiliation{%
  \institution{North Carolina State University}
  \country{USA}}
\email{trholde2@ncsu.edu}

\author{Mushfiqur Rahman}
\affiliation{%
  \institution{North Carolina State University}
  \country{USA}}
\email{mrahman7@ncsu.edu}

\author{Emily Summers}
\affiliation{%
  \institution{North Carolina State University}
  \country{USA}}
\email{emsummer@ncsu.edu}

\author{David Roberts}
\affiliation{%
  \institution{North Carolina State University}
  \country{USA}}
\email{dlrober4@ncsu.edu}

\author{Chau-Wai Wong}
\affiliation{%
  \institution{North Carolina State University}
  \country{USA}}
\email{chauwai.wong@ncsu.edu}

\author{Alper Bozkurt}
\authornote{Corresponding author: Dr. Alper Bozkurt, aybozkur@ncsu.edu}
\affiliation{%
  \institution{North Carolina State University}
  \country{USA}}
\email{aybozkur@ncsu.edu}

\renewcommand{\shortauthors}{Holder et al.}

\begin{abstract}
Animal Assisted Interventions (AAIs) involve pleasant interactions between humans and animals and can potentially benefit both types of participants. Research in this field may help to uncover universal insights about cross-species bonding, dynamic affect detection, and the influence of environmental factors on dyadic interactions. However, experiments evaluating these outcomes are limited to methodologies that are qualitative, subjective, and cumbersome due to the ergonomic challenges related to attaching sensors to the body. Current approaches in AAIs also face challenges when translating beyond controlled clinical environments or research contexts. These also often neglect the measurements from the animal throughout the interaction. Here, we present our preliminary effort toward a contact-free approach to facilitate AAI assessment via the physiological sensing of humans and canines using consumer-grade cameras. This initial effort focuses on verifying the technological feasibility of remotely sensing the heart rate signal of the human subject and the breathing rate signal of the dog subject while they are interacting. Small amounts of motion such as patting and involuntary body shaking or movement can be tolerated with our custom designed vision-based algorithms. The experimental results show that the physiological measurements obtained by our algorithms were consistent with those provided by the standard reference devices. With further validation and expansion to other physiological parameters, the presented approach offers great promise for many scenarios from the AAI research space to veterinary, surgical, and clinical applications.
\end{abstract}

\keywords{human-canine interaction, contact-free physiological sensing, computer vision, signal processing}

\maketitle

\section{INTRODUCTION}
Animal Assisted Interactions (AAIs) include a range of experiences from simple and daily interactions with pet animals to more structured canine-assisted therapies with a companion or assistance animal. AAIs are based on dynamic, mutually beneficial human-animal relationships that positively influence the health and well-being of all involved~\cite{holder2020systematicpart1, o2010companion}. With millions of dog owners around the world and hundreds of Canine Assisted Intervention programs in hospitals, the benefits of these interactions keep showing promise to cope with psychological challenges of life such as reducing emotional stress and fatigue. As such, researchers in this field are very interested in the nature of the bond across species, in understanding exactly what is occurring during various AAIs for either subject, and in the outcomes of these interactions. 

For human participants, tools such as behavior coding, psychological surveys, and biochemical analysis predominate, but these approaches face data analysis challenges as these may be anecdotal, not objective, qualitative, not real-time, not continuous, ergonomically uncomfortable, using nonstandardized devices, or dependent on complicated data collection processes~\cite{holder2020systematicpart1, ricardo2021ideation}. Similarly, some specific problems with current measurement tools for dogs mirror the issues for human monitoring in AAIs, including multispecies or body size related ergonomic issues, absence of user-focused design paradigms, difficulties measuring signals without shaving the fur to attach sensors or using subdermal insertions, and a general inability to transfer measurement or observation to naturalistic settings~\cite{foster20183d, ricardo2021ideation}. While still providing some useful information, these missing attributes limit the studies conducted in the field to go beyond generalized positive or negative human subject experiences. They also circumscribe potential answers to several detailed AAI questions about specific clinical or affective outcomes, cross-species bonding, the canine's perspective, individualizing interaction partners or protocols to increase benefit, and appropriate theoretical frameworks for interpreting these dyadic relationships \cite{holder2020systematicpart2, holder2020systematicpart1}.

Since the field of dyadic, cross-species, objective measurement is nascent, no available commercial systems provide the parameters that fully meet the aforementioned research gaps \cite{csoltova2020we}. There is a critical need for integrated systems of sensors and algorithmic tools capable of objective physiological monitoring of both AAI subjects simultaneously. Some promising evaluation techniques have started to emerge for simultaneous, non-invasive, quantitative, and objective monitoring of both human and canine subjects. For example, some progress has been made on both capacitively coupled biopotential recording systems and wearable wireless sensor systems custom-designed for animal-human dyads \cite{ahmmed2021noncontact, foster2018system}. However, despite these notable advances, these systems still need to be on the body which may bring additional ergonomic expense to researchers and subjects alike, or may be eminently susceptible to motion artifacts. These approaches still have room to grow to facilitate scalability across typical interaction environments, research contexts, and regular consumer usage. 

An interesting alternative approach is to use camera-based remote physiological monitoring methods that do not require physical contact or proprietary infrastructure for data collection. From human subjects, the seminal and subsequent research in this field has found some success in extracting skin color signals that are highly related to heart rate (HR) and blood oxygenation, in stationary or even in substantial motion use cases~\cite{li2014remote, zhu2017fitness, chen2018deepphys, mathew2021remote, tian2022multi}. In this vein, extensive human remote physiological monitoring work has been done in the literature to detect HR in physical exercising contexts~\cite{zhu2017fitness} and extracting a blood oxygenation signal from the hand~\cite{mathew2021remote, tian2022multi}. The animal remote physiological monitoring landscape is much more limited with most work being done in dogs using radar or infrared devices rather than regular red-green-blue (RGB) cameras \cite{rizzo2017monitoring, wang2019distinction, wang2020non, wang2019method}, or in computer vision detection of physiological signals in cattle, pigs, and exotic animals \cite{al2019pilot, maria2019computer, jorquera2019modelling, jukan2017smart, barbosa2019contactless, pereira2019perspective, wang2021contactless}.

Here, we explore camera-based remote physiological sensing for monitoring both interactants simultaneously---detecting human HR and canine breathing rate (BR) simultaneously. To the best of our knowledge, this is the first time these two efforts are reported together in the literature. This paper presents a proof-of-concept work by, first, providing the details of the \textit{in vivo} experimental setup and the signal processing algorithmic requirements for remote physiological monitoring in AAIs, and then comparing the derived remotely monitored outputs to those of reference wearable physiological devices and human observation analyses. 

While providing data relevant to both participants' psychophysiological states, these novel approaches bring the potential of removing the need for shaving animal fur, attaching sensors to the body, straps, harnesses, and other experimental accouterments to assess the physiological state during the interaction \cite{becker2006fundamentals, singh2018heart}. This work also holds the potential to be deployed unobtrusively in most domestic environments, to reduce or eliminate the Hawthorne effect in AAI research, and to be a useful tool in veterinary surgical, tele-veterinary monitoring, and animal shelter scenarios. Finally, noting the paucity of remote physiological monitoring research in non-human animals, this work paves the way for further exploration of these camera-based physiological monitoring approaches in other non-human animals generally and in more human-animal interactions specifically.

\section{METHODS}
In order to validate the accuracy of the aforementioned camera-based system for simultaneous estimation of such physiological data as human HR and canine BR, we conducted an \textit{in vivo} proof-of-concept study. All experimental procedures were approved by the Institutional Review Board for the Use of Human Subjects in Research and the Institutional Animal Care and Use Committee of the authors’ institution. Specifically, our AAI remote physiological assessment setup included standard reference wearable devices, multiple cameras, significant back-end signal processing, and relevant analytics. 

\subsection{Participants}
For this proof-of-concept case study, the subjects were an adult female human and an adult male human individually interacting with either a 10-year-old male dog subject (Shih-Tzu) or a 5-year-old female dog subject (Pembroke Welsh Corgi). 

\subsection{Research Setup}
The tests were conducted in a clear, demarcated lab space to simulate a typical AAI scenario. The main camera was set up at the end of the table with one or two auxiliary cameras on either side of the interacting dyad, so that the human subject could sit up comfortably while also interacting with the dog subject.

\subsection{Procedure}
Before the tests, the human subjects were debriefed on the study protocol and screened for COVID-19. All reference devices and cameras were then synchronized using both a software event marker and a physical synchronization event. The human subject then donned the reference wearable devices while a researcher affixed the wearable reference devices to the dog. With the researcher in the corner of the room, the human and canine subjects proceeded to interact calmly within the demarcated research space for 10 minutes before a brief rest period followed by another 10-minute interaction. The canine subject was free to leave the interaction at any time and appropriate  rewards were anticipated to come from the pleasant patting during the interaction, from positive words spoken by the human interactant, and additional treats to the canine subject. Researchers were careful to note and appropriately respond to either subjects' signs of discomfort with the interaction, reference devices, or any other elements of the research protocol.

\subsection{Reference Acquisition}
\begin{figure}[!t]
\centering
\includegraphics[width=\linewidth,trim={0.2cm 0cm 0.1cm 0cm},clip]{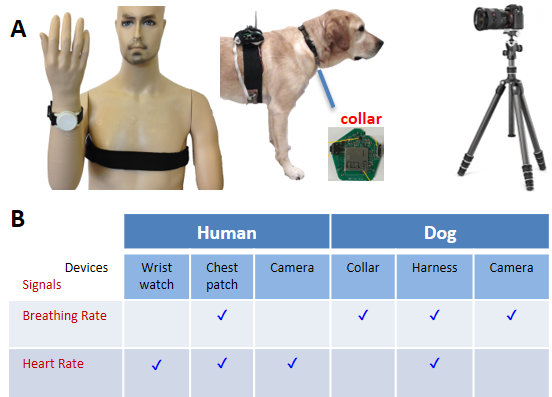}
\caption{Camera-based setup for remote HR and BR acquisition and wearable systems for reference signal measurement.
(A) The custom wearable and noncontact (camera) devices used in the experiments. (B) Summary of the signals received from each device.}
\label{fig:fig1}
\Description{Two photos labelled (a) and (b) show a human and a canine wearing some physiological measurement devices, a camera setup and a table portraying the relevance of the devices with the body functions. Image (a) shows a human wearing a wristband and a chest patch, a canine with a smart collar and a vest on its neck and torso and a digital camera hold on a tripod. Image (b) depicts a table with tick marks representing the relevance of each device with heart rate and breathing rate for human and canine separately.}
\end{figure}

For human subjects, we used a commercially available HR detection device (E4 wristband, Empatica Inc., MA, USA) as reference during the experiment \cite{schuurmans2020validity}. The other reference systems were custom-designed wearable wireless sensing platforms including a vest and smart collar for dogs and a chest patch and another wristband for humans to be worn throughout the interaction [Figure ~\ref{fig:fig1}(A)] \cite{dieffenderfer2016low, foster2019preliminary, williams2020smart}. The second wrist subsystem for humans monitors photoplethysmography (PPG) based HR and the movement of the dominant arm. The chest subsystem collects electrocardiogram (ECG) based HR, PPG based HR, and motion physiological data. This chest subsystem and accompanying electrodes are attached to the torso by medical-grade tape and endogenous adhesive, respectively. The canine reference devices measure HR---via ECG with custom-designed 3D-printed electrodes that bypass animal fur---and collect movement from the neck and torso body locations. Human visual observation is also used to obtain reference signal of canine BR from the main camera video. For our task, this combination of reference modalities provides redundancy that allows for cross validation.

\subsection{Camera Specifications}
Commercially available, low-cost camera systems were utilized throughout this research. This noncontact system consists of high-resolution RGB cameras [Figure ~\ref{fig:fig1}(A)] placed within the evaluation space's perimeter (Figure ~\ref{fig:fig1}, embedded table for measured data). A 0.3-megapixel, HD fixed focus camera situated in the bezel of a Lenovo Flex 3 laptop (Lenovo Group Limited, China) served as the main lens for these experiments \cite{lenevoflex3specs, lenevoflex3}. 

The first auxiliary system was the rear-facing cameras of a Galaxy S10 phone (Samsung Electronics, South Korea) featuring a 12-megapixel telephoto lens at 45$^\circ$, a 12-megapixel wide-angle lens at 77$^\circ$, and a 16-megapixel ultra-wide-angle lens at 123$^\circ$ \cite{galaxys10specs}. 

The second auxiliary camera was the rear-facing, 12-megapixel, autofocusing camera of an iPhone 6s (Apple Inc., CA, USA) \cite{iphone6sspecs}.

\subsection{Physiological Signal Processing}
\subsubsection{HR and BR Estimation}
\begin{figure}[!t]
\centering
\includegraphics[width=\linewidth,trim={1.6cm 0.8cm 1.6cm 0.8cm},clip]{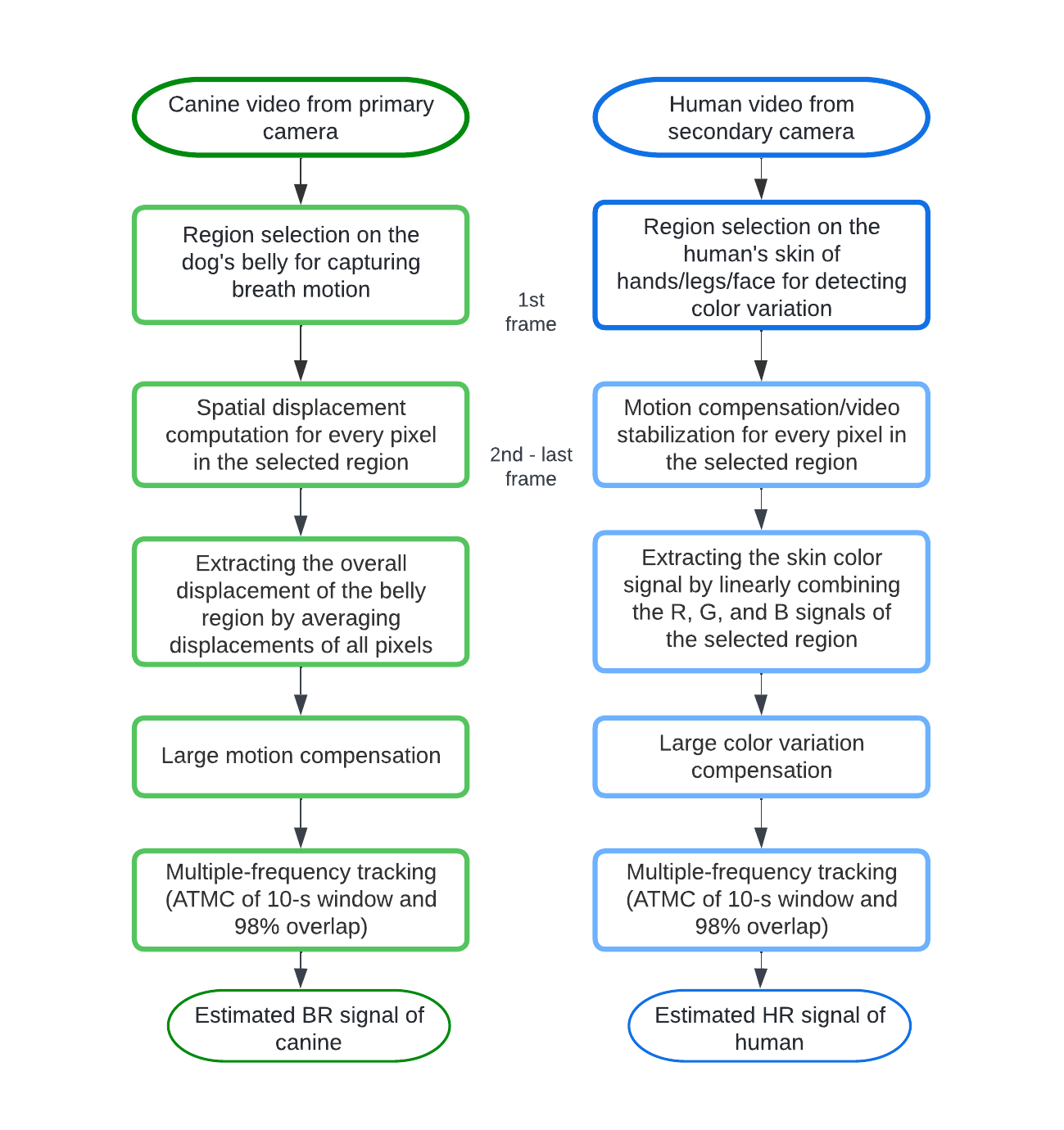}
\caption{Flowchart of the proposed processing methods for human HR and canine BR estimation. The sequential pipelines separate the breath/skin color signal from the overall video signal in the time domain and later in the frequency domain.}
\label{fig:fig2}
\Description{Two process flow diagrams aligned vertically side by side. The left diagram starts from 'Canine video from primary camera' and ends at 'Estimated BR signal of canine' with five intermediate stages. The right diagram starts from 'Human video from secondary camera' and ends at 'Estimated HR signal of human' with five intermediate stages.}
\end{figure}

Our approach of contact-free physiological sensing in this paper focuses on processing the video streams of a canine and a human to extract the time-series signals of HR and BR, respectively. By applying motion estimation to a human/canine video, our approach can generate an intermediate, stabilized video to facilitate the precise extraction of physiological signals. Motion estimation allows tracking a specific point on a human’s skin or a canine’s body/fur even though the location of the point in the captured video stream may drift over time. We pick different visual features from the human and canine videos to estimate HR and BR signals. For canine videos, we focus on the subtle periodic movement of the canine’s belly due to breathing. The overall displacements of the pixels around the belly are aggregated to obtain a periodic signal reflecting the breath. For human videos, we focus on extracting the subtle, periodic variation of the color of the human skin as a result of the periodic blood volume change induced by the heartbeat \cite{zhu2017fitness}. An intermediate skin color signal is subsequently processed via a robust frequency tracker to extract the HR time-series signal. For both video types, the periodic signals of our interest may have a much smaller amplitude than other dominating signals such as those caused by large/abrupt motion of human/dog, camera tilting, and illumination change of light sources. The adaptive multi-trace carving (AMTC) algorithm \cite{zhu2020adaptive} is employed to unveil the periodic micro-signals of interest. Flowcharts of our proposed methods for HR and BR estimation are shown in Figure ~\ref{fig:fig2}, respectively.

\subsubsection{Motion Estimation Over Video Frames}
Accurate pixel-based motion estimation over the video frames is necessary to unveil the subtly embedded micro-variations. Given two video frames, optical flow algorithms can compute pixelwise motion as flow vectors \cite{liu2009beyond}. For canine videos, the result of optical flow motion estimation is used in two ways. First, in our recorded videos, the pixel locations of the canines’ belly shift within the video frames due to the canines’ natural tendency to wander or to have a more comfortable body posture while interacting with humans. Optical flow is used for tracking the new position of the dog’s belly in the target frame by exploiting the gradient information between the first frame with known pixel locations and the target frame with unknown pixel locations. Second, the spatial average of the motion vectors over the belly region associated with each frame is further used for extracting the time-series signal that contains subtle breathing motion. On the other hand, for HR estimation from human videos, the spatial motions with respect to the first frame of the video sequence are estimated to obtain the new pixel locations of the skin region in the target frame.

Previous work on contact-free physiological sensing of human HR has used optical flow estimation for fine-grain alignment of the cheek region \cite{zhu2017fitness}. In the work we present here, we evaluated both conventional algorithms \cite{liu2009beyond} and deep-learning-based algorithms such as the recurrent all-pairs field transforms~(RAFT)~\cite{teed2020raft} and the global motion aggregation~(GMA)~\cite{jiang2021learning} for optical flow estimation. We found that the GMA performs the best throughout the experiments of human-canine interaction. GMA is based on RAFT, which adopts an iterative approach for flow field estimation through recurrent neural networks \cite{teed2020raft}. RAFT builds a multi-scale correlation volume for each pair of pixels by extracting pixel features. GMA uses an additional block of image self-similarity-based global motion aggregation to handle occlusion cases \cite{jiang2021learning}. The work presented in \cite{zhu2017fitness} divided each video into small segments and used pre-alignment of the cheek region for each small segment to avoid occlusion during optical flow estimation. This required to preset the boundary of the cheek region every 1.5 seconds for smooth optical flow estimation. In contrast, in this paper, we use pre-alignment of the canine belly or human hands only once in the first frame of each video. It, then, smoothly computes optical flow throughout the video with an average length of 132 seconds. As another benefit, this eliminated the need of tackling signal discontinuation issue while concatenating the small segments \cite{zhu2017fitness}.

Optical flow estimation requires a reference frame and a target frame from a video sequence for calculating the gradient. In this work, we estimated optical flow from two types of reference frames---the first frame of the video and the previous frame immediately before the current one. We found out that the latter approach results in a significant error as we temporally accumulate the optical flow vectors to obtain the pixel displacement with respect to the first frame. We, therefore, used the first frame as the reference for optical flow estimation in this study.

\subsubsection{Region Selection and Remote Physiological Signal Extraction}
\begin{figure*}[!t]
\centering
\includegraphics[width=\linewidth, trim={0cm 0cm 0cm 0cm},clip]{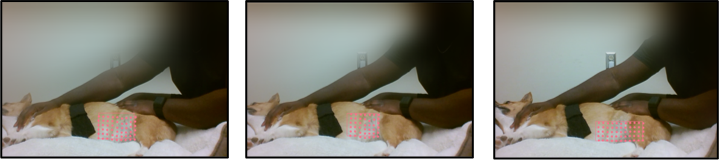}
\caption{Example of three independently chosen rectangular regions (shown in matrices of red dots) on the canine’s belly for collecting breath motion. The regions are large enough so that the effect of local motion or noise is reduced.}
\label{fig:fig6}
\Description{Three vertically aligned pictures each picture showing a dog lying on the ground and a human's hand hold on the dog's neck. On top of that grid points on a rectangular shape are illustrated on the dog's belly locating at different pixel positions on the three pictures.}
\end{figure*}

For the canine videos, we draw a rectangular bounding box inside the belly region for extracting micro-level breathing signals (Figure~\ref{fig:fig6}). We can extract certain properties (i.e., color, location, motion) of the selected pixels from each video frame to obtain a time-series signal. As the canine breathes, a periodic motion is observed in this region along the vertical axis of the dog’s belly. In our collected videos, this vertical axis roughly aligns with the $y$-axis of the camera frame. For this reason, we use the optical flow vector along the $y$-axis as the source signal in our experiment. We, then, take the spatial average of the motion vector for all selected pixels of the belly to reduce the effect of local noise. Note that, the locations of selected pixels get updated in each frame according to optical flow-based motion estimation. After taking the motion vector average in each of the frames, we obtain a time series which is the remote physiological monitoring signal of the dog and contains the dog's breathing information.

Human videos are processed similarly for updating pixel locations starting from the selected skin region in the first frame of the video. For human HR estimation, we take the color values of red, green, and blue channels and for each channel, we take the average across all the selected pixels. The signal of heartbeat is buried in the RGB color values along with other sources of variation such as face motion \cite{zhu2017fitness}. We convert the three-channel time sequence signal into a one-dimensional signal by using balanced weights $(1,1,1)$ for the channels. We also tried linear weights $(-1,2,-1)$ as in \cite{zhu2017fitness}, but we found $(1,1,1)$ performs better. This extracted signal is the remote physiological monitoring signal of humans and corresponds to the human HR.

While the canine belly is the suitable region for extracting the remote physiological monitoring signal related to canine BR, the remote physiological monitoring signal related to human BR can be extracted from face, legs, or hands. The selected region should be large enough to avoid noise originating at the camera sensor (for human HR extraction) or artifacts found in the motion vector estimation output (for canine BR extraction). For statistical verification, for each video, we tested our algorithm on three rectangular selected regions independently drawn inside the dog’s belly or human’s face, hands, or legs.

\subsubsection{Refining Subtle HR/BR Signals}
The extracted remote physiological monitoring signal corresponding to human HR or canine BR is submerged within signals from multiple other sources, where the sources related to human or canine’s wandering motion has a dominance. To decouple the subtle breath motion or heartbeat signal from the mixture of the signals, we go through several signal processing steps starting from the remote physiological monitoring signal.

\begin{figure*}[!t]
\centering
\includegraphics[width=1\linewidth, trim={2.6cm, 0.8cm, 3.2cm, 1.4cm}, clip]{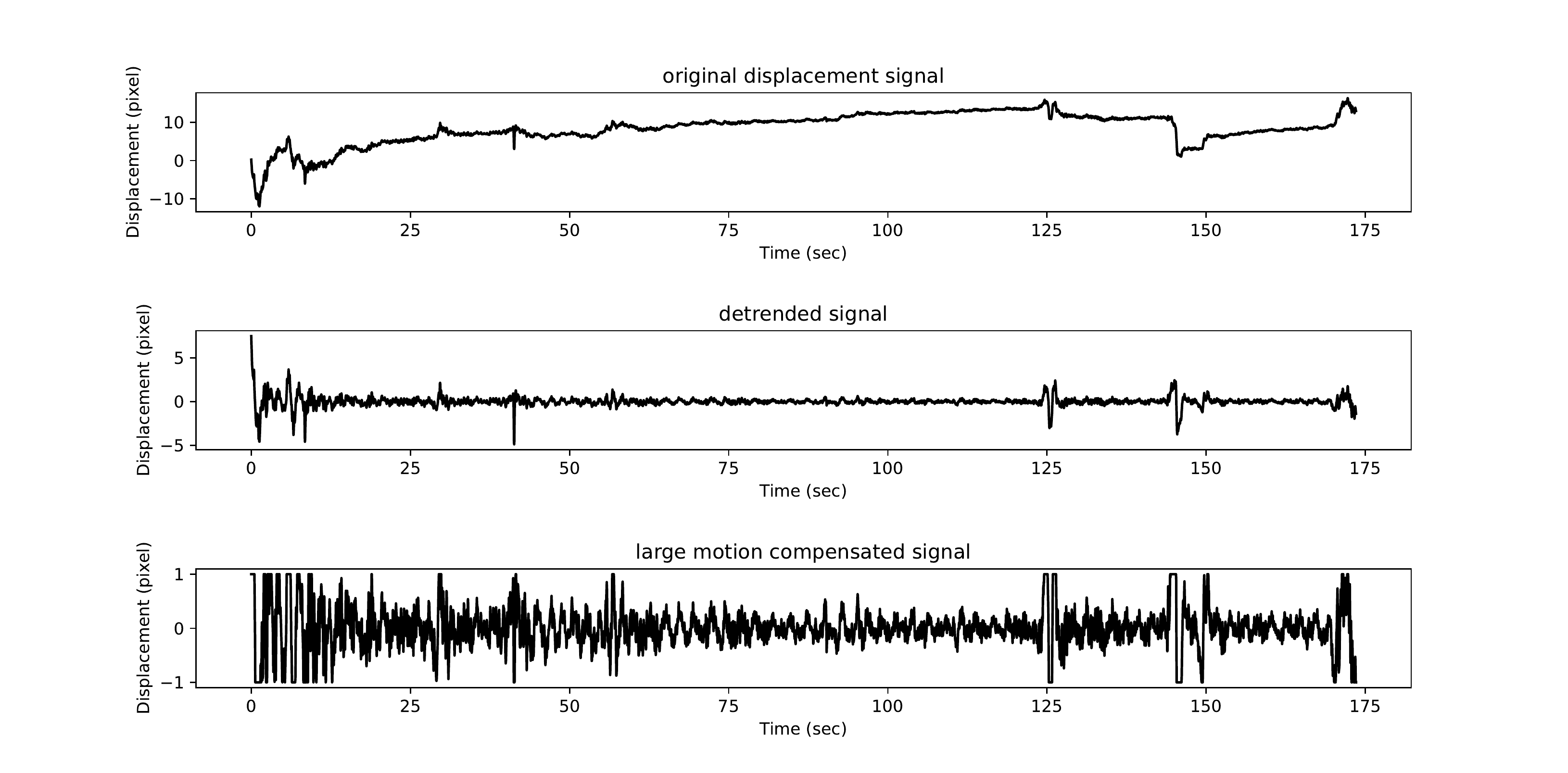}
\caption{Representative outcome of the signal processing steps for refining subtle breath motion. Periodic nature of the breath motion signal is perceivable only after detrending/motion stabilization. Breath motion does not exceed a small range.}
\label{fig:fig3}
\Description{Three horizontally aligned graphs. The topmost graph shows original displacement signal from -10 to 10 pixels on the Y axis against time from 0 to 175 seconds on the X axis. The centered graph shows detrended displacement signal from -5 to 5 pixels on the Y axis against time from 0 to 175 seconds on the X axis. The bottom graph shows large motion compenstaed displacement signal from -1 to 1 pixels on the Y axis against time from 0 to 175 seconds on the X axis.}
\end{figure*}

\begin{figure*}[!t]
\centering
\includegraphics[width=1\linewidth, trim={2.5cm, 0.8cm, 3.2cm, 1.0cm}, clip]{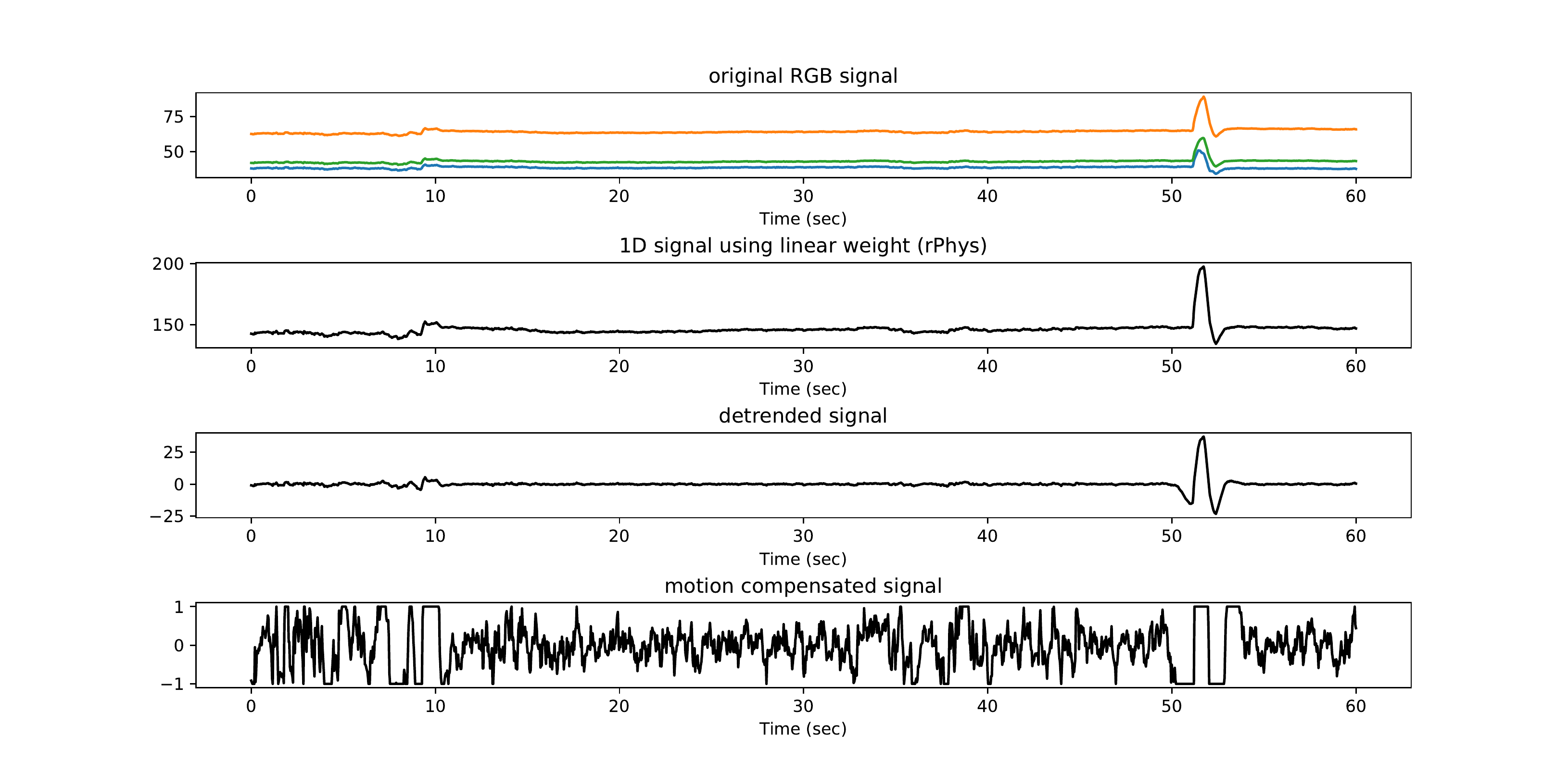}
\caption{Representative outcome of the signal processing for refining the color variation caused by heartbeat.  A periodic variation of the linearly combined color signal is found within a small range after removing the drastic variations.}
\label{fig:fig4}
\Description{Four horizontally aligned graphs. The topmost graph shows original RGB signal containing three colored lines representing red, green and blue channels on the Y axis against time from 0 to 60 seconds on the X axis. The next graph shows remote physiological monitoring signal on the Y axis against time from 0 to 60 seconds on the X axis. The next graph shows detrended remote physiological monitoring signal on the Y axis against time from 0 to 60 seconds on the X axis. The last graph shows large motion compenstaed signal on the Y axis against time from 0 to 60 seconds on the X axis}
\end{figure*}

First, we estimate the trend of the extracted signal by calculating a moving average of suitable sample length ($\sim\!60$ samples). Then, we subtract the baseline/trend from the original signal to obtain the detrended signal. However, these detrended signals still have some sharp variations due to the canine's abrupt movement or specular reflection on the skin when the human subject moves. The detrended signal can be clipped to remove the effect of large movement. We clip the motion signal at 1 pixel, which we considered as the maximum average displacement originating from the subtle breath. Similarly, we clipped the detrended color signal at 1 out of 256 shades of gray for human HR estimation. Figs.~\ref{fig:fig3} and~\ref{fig:fig4} illustrate the processing of the canine BR signal and human HR signal, respectively.

We, then, further refine the time domain signal by taking small segments of about 2~seconds and passing each of those through a process of standardization using zero mean and unit variance. We take overlapping small segments in this step starting from every sample in the sequence that provides a valid window. Finally, we accumulate the processed segments through averaging at each temporal location.

\subsubsection{Frequency Estimation and Tracking}
\begin{figure}[!t]
\centering
\includegraphics[width=\linewidth, trim={0.8cm 0.2cm 1.6cm 0.4cm},clip]{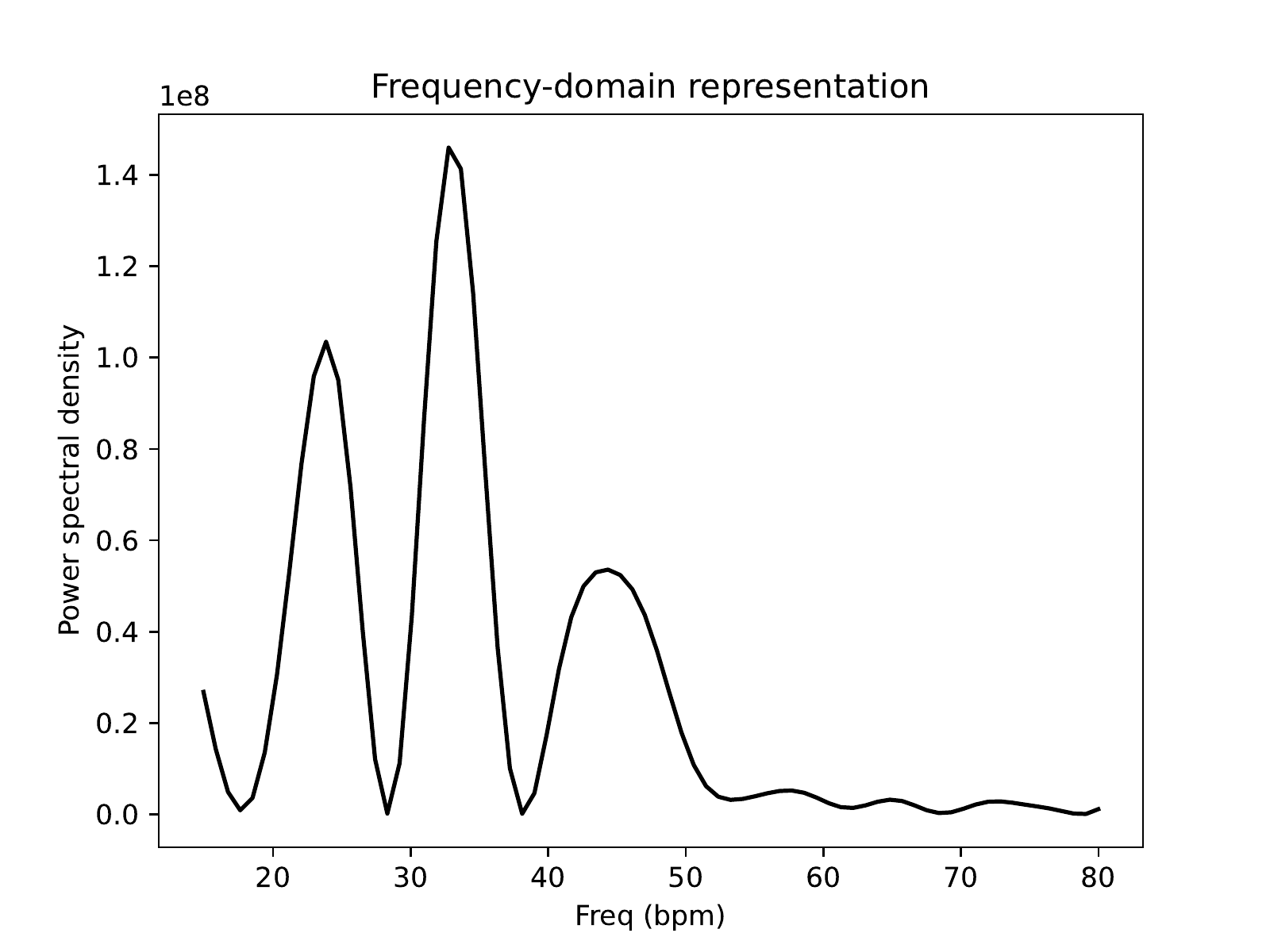}
\vspace{-3mm}
\caption{Representative estimated power spectral density of a processed breathing motion signal of 10~seconds. Within the possible range of breath rate (15--50 bpm), multiple frequencies pose peaks. When multiple consecutive temporal segments are put together in a spectrogram, AMTC can be used for robust frequency tracking by exploiting the temporal smoothness of the frequency signal.}
\label{fig:fig5}
\Description{A graph shows power spectral density on the Y axis against frequency from 10 to 80 bpm on the X axis.}
\end{figure}

Fast Fourier transform (FFT) of the remote physiological monitoring signal is computed between a start time and end time to obtain the spectral density for a probable frequency range. Figure ~\ref{fig:fig5} shows an estimated power spectral density of a sample processed window signal of 10~seconds. It is evident from the plot that there are multiple sources of frequencies present in the extracted remote physiological monitoring signal. Choosing the frequency of the peak may not reflect the desired frequency. Within this small window, other sources of frequency may be comparable or even stronger than the physiological source. Hence, an algorithm is needed to jointly estimate the power spectral density functions at various temporal locations and iteratively compute the best time-series of HR/BR frequencies.

We used the adaptive multi-trace carving (AMTC) algorithm~\cite{zhu2018adaptiveconf, zhu2020adaptive} for robust frequency estimation and tracking for both canine BR and human HR estimation. AMTC can track frequency traces in a spectrogram by taking into account the smoothness of the trace~\cite{zhu2020adaptive}. A dynamic programming based approach is used to estimate the current frequency at every time step and update the frequency values of previous time steps through backtracking. AMTC can take many parameters including window length, percentage of overlap, the number of FFT points, and backtracking length. In this effort, we used a 10-second window, 98\% overlap, 2048 points of FFT, and 10 samples of backtracking length. Optionally, AMTC is also capable of tracking multiple frequency traces in an iterative manner, if there are other frequency traces available than HR or BR. In this way, spectrograms and frequency traces of canine BR and human HR are obtained from AMTC.

\section{RESULTS}
The described physiological signal processing method in the previous section was tested on the videos collected in our \textit{in vivo} proof-of-concept study. To compare the human HR results with standard reference devices, we used the data provided by the E4 (Empatica Inc., MA, USA) wristband software.  For comparing the canine BR to an independent reference, we used both the outcome of the wireless wearable sensor and trained a human operator to observe and annotate dog breathing events carefully from the videos. We used root mean squared error (RMSE) and relative error rate (MeRate) as the standard metrics \cite{zhu2017fitness} to compare remote physiological measurements with the standard reference devices in the context of AAI.

\begin{figure*}[!t]
\centering
\includegraphics[width=\linewidth]{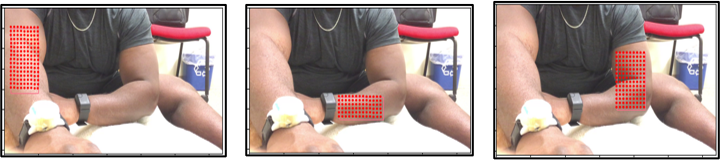}
\caption{Example of independently drawn rectangles (shown in matrices of red dots) on the human skin for collecting periodically varying color signals. Rectangles are large enough to suppress locally originated color bias due to either specular reflection or insufficient lighting.}
\label{fig:fig7}
\Description{Three vertically aligned pictures each picture showing a human with visible skin of two hands and one leg. On top of each picture, grid points on a rectangular shape are illustrated on either of the human's hands or the leg.}
\end{figure*}

\begin{figure*}[!t]
\centering
\includegraphics[width=\linewidth,trim={0cm 0cm 0cm 0cm},clip]{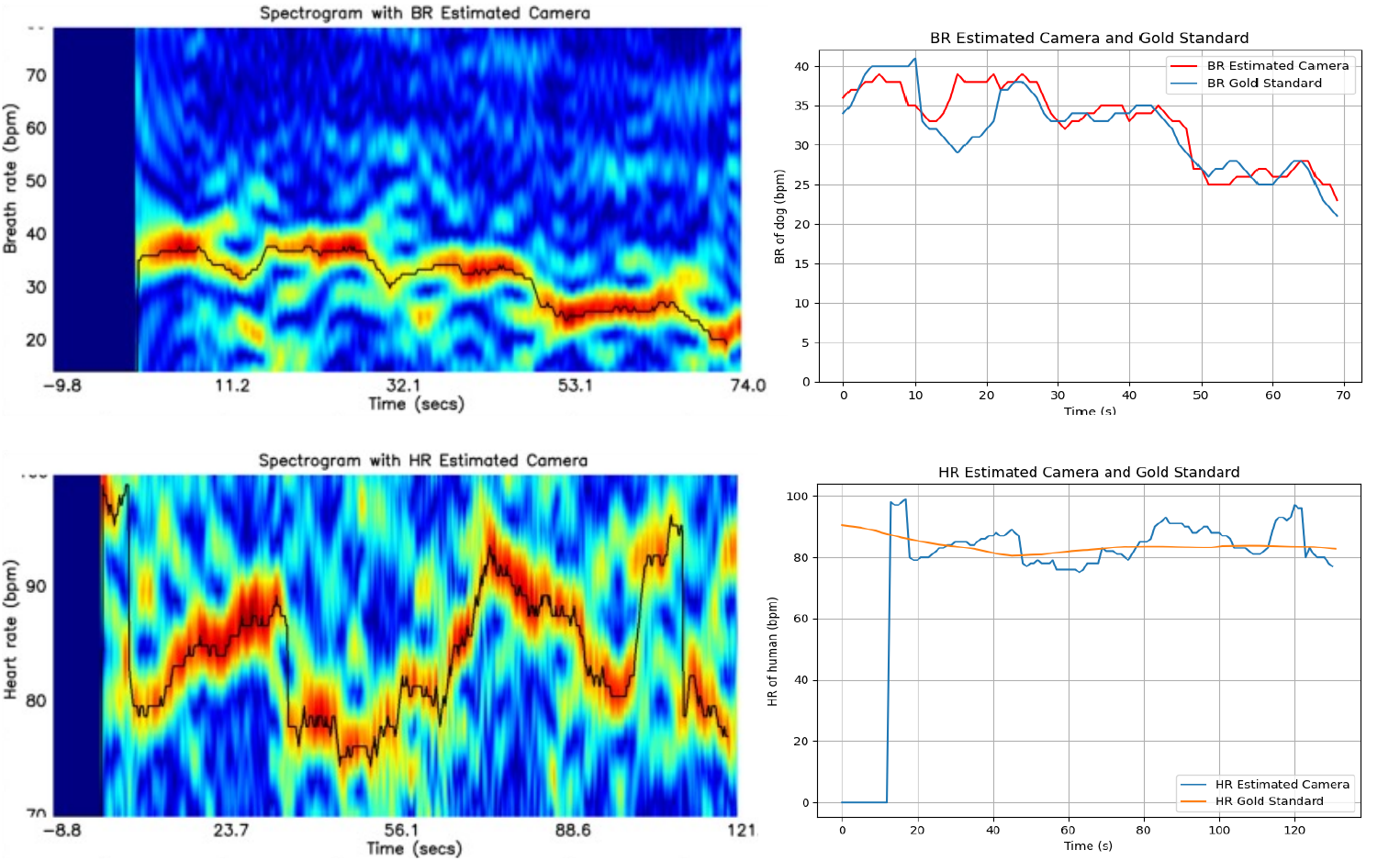}
\vspace{-5mm}
\caption{(Top left) Spectrogram obtained from the processed breathing motion signal with an overlaid frequency trace produced by AMTC for the case-study canine video analyzed. (Top right) AMTC estimated canine BR and reference canine BR calculated from the video. (Bottom left) Spectrogram obtained from the processed color signal with an overlaid frequency trace produced by AMTC for the human subject. (Bottom right) AMTC estimated human HR and reference human HR calculated from the video.}
\label{fig:fig8}
\Description{Four subplots on a two by two square grid showing spectrogram related with BR or HR estimation and the comparison with estimated HR or BR with the reference HR or BR on the same plot. The top left subplot shows the spectrogram of BR estimation of 10 to 80 bpm in the Y axis and -9.8 to 74 seconds on the X axis. The top right subplot illustrates estimated BR and reference BR of dog on the same plot having BR from 0 to 40 bpm on the Y axis and time from 0 to 70 seconds on the X axis. The two plots almost match with each other with a noticeable disparity from 10 to 20 seconds. The bottom left subplot shows the spectrogram of HR estimation of 70 to 100 bpm in the Y axis and -8.8 to 121 seconds on the X axis. The bottom right subplot illustrates estimated HR and reference HR of human on the same plot having HR from 0 to 100 bpm on the Y axis and time from 0 to 130 seconds on the X axis. The plot of estimated HR oscillates centering the reference HR.}
\end{figure*}

We evaluated the results by comparing estimated BR from three dog videos and HR from four human videos with the corresponding ground truth. For each video, we ran our algorithm three times using three sets of rectangular regions drawn independently. Figs.~\ref{fig:fig6} and \ref{fig:fig7} show two sets of representative selected regions for one dog video and one human video, and Figure ~\ref{fig:fig8} illustrates the resulting spectrograms and estimated canine BR and human HR for two representative videos using AMTC. Table~\ref{table:table1} displays the resulting accuracy metrics for these two showcased videos.

\begin{figure*}[!t]
\centering
\includegraphics[width=\linewidth,trim={0cm 0cm 0cm 0cm},clip]{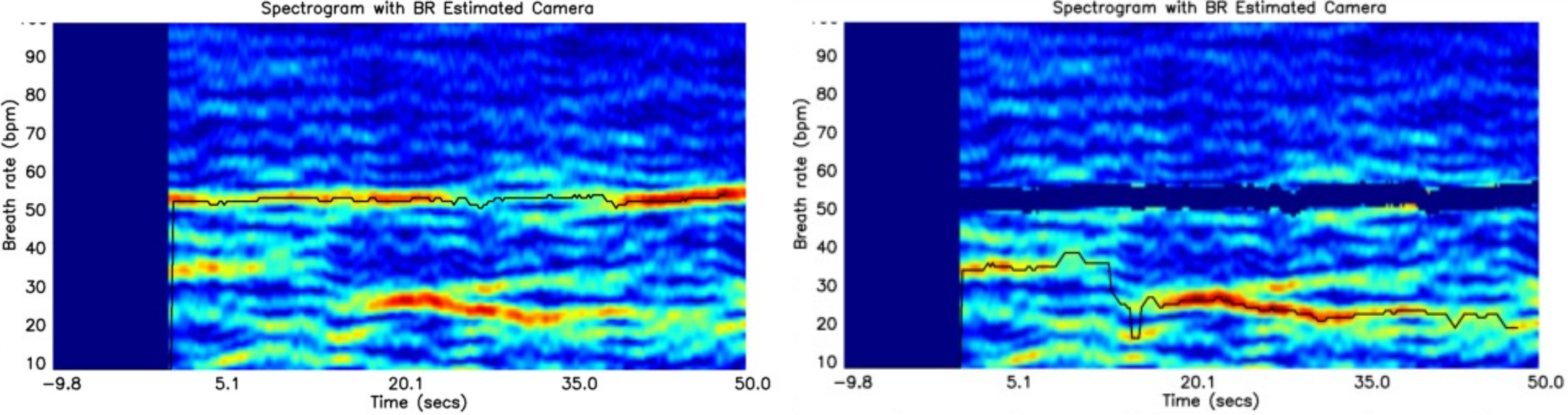}
\vspace{-5mm}
\caption{(Left) Spectrogram and detected trace at the first internal iteration of AMTC. (Right) Spectrogram and detected trace at the second internal iteration of AMTC. The first and second frequency traces correspond to the patting intensity and canine breath rate respectively, both detected from the selected region of the dog’s belly.}
\label{fig:fig9}
\Description{Two vertically aligned plots showing two spectrograms related with BR estimation. The left plot shows a spectrogram of 10 to 100 bpm in the $y$-axis and -9.8 to 50 seconds on the $x$-axis. A trace line is illustrated on top of this spectrogram between 50 and 60 bpm. The right plot shows a spectrogram of 10 to 100 bpm in the $y$-axis and -9.8 to 50 seconds on the $x$-axis. A trace line is illustrated on top of this spectrogram between 20 and 40 bpm.}
\end{figure*}

\begin{table}[!t]
  \caption{Performance of HR and BR estimation method on two case-study videos.}
  \vspace{-3mm}
  \label{table:table1}
  \begin{tabular}{cccc}
    \toprule
    Task                    & RMSE& SD|error|& MeRate \\ 
   &(bpm) & (bpm) & \\ 
    \midrule
    BR estimation of canine & 2.9        & 2.1             & 6.4\% \\
    HR estimation of human  & 5.2        & 2.9             & 5.1\% \\ 
    \bottomrule
  \end{tabular}
\end{table}

	As proof of concept, Table~\ref{table:table2} and Table~\ref{table:table3} show the overall evaluation results of the used contactless HR and BR estimation methods across the recorded videos using standard metrics. We compare the results by disabling one or two signal processing techniques (i.e., detrending, standardization, and clipping) that we use along with AMTC. The best performance is obtained as a 6.5\% relative error for BR estimation from dog videos and a 9.7\% relative error for HR estimation from human videos. The required approach for this result occurs when the extracted physiological signal is first detrended using a 2-second window, then clipped between -1 and 1 before standardization using either a 2-second (for HR) or a 4-second (for BR) window. Overall, these initial results demonstrate the promise of the used RGB camera-based signal processing approach to extract human HR and canine BR simultaneously during a human-canine interaction.

\begin{table}[!t]
  \caption{Evaluation of method for human HR estimation.} 
  \vspace{-2mm}
  \label{table:table2}
  \begin{tabular}{cccc}
    \toprule
    Contact-less HR est.& RMSE & SD|error|& MeRate         \\
    setting (Human) &(bpm)   & (bpm) & \\
    \midrule
    Standardization                                  & 8.0          & 4.7             & 7.4\%          \\
    Detrending +   Standardization                   & 6.9          & 4.1             & 6.5\%          \\
    Detrending + Clipping + & 7.0 & 4.2    & 6.5\% \\ 
    Standardization & & \\
    \bottomrule
  \end{tabular}
\end{table}

\begin{table}[!t]
  \caption{Evaluation of method for canine BR estimation.} 
  \vspace{-2mm}
  \label{table:table3}
  \begin{tabular}{cccc}
    \toprule
    Contact-less BR est.& RMSE& SD|error|& MeRate         \\
     setting (Canine)          & (bpm)   & (bpm) &  \\
    \midrule
    Standardization                                  & 8.0          & 5.2             & 11.2\%         \\
    Detrending +   Standardization                   & 6.1          & 4.7             & 9.7\%          \\
    Detrending + Clipping + & 6.2 & 4.9    & 9.7\% \\ 
    Standardization & & \\
    \bottomrule
  \end{tabular}
\end{table}

	As an exploratory analysis, we also ran AMTC on a special video containing the case of periodic patting behavior of the human hand during AAI. This was to test the possibility of detecting canine breath motion when the breath signal is not the primary source of overall periodic motion. As shown in Figure ~\ref{fig:fig9}, at the first internal iteration, the AMTC returns a trace that resembles the trace of the human hand patting rate. At the second internal iteration of AMTC, another trace is observed and empirically determined to be the trace of canine breathing rate. Human hand patting is expected to provide a strong periodic motion along the horizontal direction on the canine's body. As a future extension of this work, remote physiological signals from both $x$- and $y$-axes can be used to automatically recognize canine BR and patting frequency traces. This additional result has implications for noncontact, non-observer tracking of interaction intensity and frequency, both being metrics of interest to the AAI community \cite{holder2020systematicpart2}.

\section{DISCUSSION}
This preliminary study helped us to explore several challenges when estimating human HR and dog BR from videos. Initially, we also tested the idea of estimating canine HR (in parallel to BR) by extracting skin color similar to human HR analysis. This was naturally difficult due to the dog's fur where shaving a window may be an effective but inconvenient solution. We ran our human HR estimation algorithm on a video containing canine's ear where the skin was roughly visible. The result was somewhat inferior compared to the human HR estimation result, which is reasonable considering the vascularization of this region and the nonstandard skin features of the canine. On the other hand, detecting the BR of a human by motion estimation, as it was done on the canine, was also more difficult due to relevant parts of the body (such as chest and its motion with breathing) being obscured by clothing. For these reasons, we chose only to verify the possibility of simultaneous contactless detection of canine BR and human HR in this preliminary work. Our future work will investigate the additional camera angles targeting the part of the dog without hair (e.g., ears) and using body-conforming clothing for human subjects to obtain HR and BR both on humans and animals. 

Another challenge is that, both the human and the animal may periodically move or change posture during AAIs, which acts as an additional source of frequency component potentially contaminating the Fourier analysis involved in remote physiological signal processing. AMTC opens the possibility of tracking multiple frequency traces and of isolating only the movement components of interest (i.e., breathing, stroking, etc.) 

The reference HR signals obtained from standard commercial devices (e.g., Empatica's E4) generally have proprietary detection algorithms leading to certain delays that differ from the time lags of our AMTC approach. For this reason, we used correlation-based matching for time synchronization of the two signals, and note that our future investigation will focus on the impact of this on our error profiles. 

Human HR estimation depends on cameras detecting skin color requiring body regions being exposed during AAIs. The quality and intensity of the lighting system in the interaction room, the direction of the light source, the angle of the skin surface with respect to the light, and inherent camera sensor noise are all parameters that affect the performance. These should be considered when selecting AAI experimental context and optimized to provide lower signal estimation errors across the board. 

Beyond these future technical improvements, the promising results in this work are very encouraging to perform larger studies and more physically active experiments to validate the techniques and assess performance during various AAI contexts and use cases. Furthermore, analysis of ocular surface temperature and other noncontact signals could be added to the analysis for a more complete package of remote physiological signal analyses relevant to AAIs \cite{csoltova2017behavioral, csoltova2020we}. Beyond AAIs, these techniques also can be extended to other human-canine interaction applications from therapy or guide dog evaluation, to the clinical or surgical veterinary context, and dog shelters. Additionally, this is also a first step toward enabling veterinary telemedicine or even remote pet therapy, especially if extended to other therapy and domestic animals \cite{lee2006mobile, best2022veterinaryservices}.

\section{CONCLUSION}
In this paper, we have presented a contact-free approach to facilitate AAI studies via physiological sensing of humans and canines using consumer-grade cameras. In this preliminary study, we remotely sensed the HR signal of human subjects and the BR signal of dog subjects while they interacted with each other. Comparing the estimates produced by the proposed methods here with the reference gold-standard measurements has shown that our approach can tolerate specific amounts of motion such as involuntary body movement. Furthermore, spurious periodic signals such as those due to human patting can be isolated to unveil the desired canine breathing signal. This proof-of-concept work presents, for the first time, an effort toward 1) remote camera-based assessment of BR in dogs, and 2) simultaneous detection of dog BR and human HR during AAIs. Further development and validation of this remote physiological sensing setup may enable free monitoring of dyadic interactions between humans and animals in a noninvasive and unencumbered way for improving the efficiency and effectiveness of various AAI applications.

\begin{acks}
The authors acknowledge the support from IBM Faculty Awards and from National Science Foundation through CCSS-1554367 and ECC-1160483. 
\end{acks}

\balance
\bibliographystyle{ACM-Reference-Format}
\bibliography{aci.bib}


\begin{thebibliography}{42}


\ifx \showCODEN    \undefined \def \showCODEN     #1{\unskip}     \fi
\ifx \showDOI      \undefined \def \showDOI       #1{#1}\fi
\ifx \showISBNx    \undefined \def \showISBNx     #1{\unskip}     \fi
\ifx \showISBNxiii \undefined \def \showISBNxiii  #1{\unskip}     \fi
\ifx \showISSN     \undefined \def \showISSN      #1{\unskip}     \fi
\ifx \showLCCN     \undefined \def \showLCCN      #1{\unskip}     \fi
\ifx \shownote     \undefined \def \shownote      #1{#1}          \fi
\ifx \showarticletitle \undefined \def \showarticletitle #1{#1}   \fi
\ifx \showURL      \undefined \def \showURL       {\relax}        \fi
\providecommand\bibfield[2]{#2}
\providecommand\bibinfo[2]{#2}
\providecommand\natexlab[1]{#1}
\providecommand\showeprint[2][]{arXiv:#2}

\bibitem[Ahmmed et~al\mbox{.}(2021)]%
        {ahmmed2021noncontact}
\bibfield{author}{\bibinfo{person}{Parvez Ahmmed}, \bibinfo{person}{Timothy
  Holder}, \bibinfo{person}{Marc Foster}, \bibinfo{person}{Ivan~D Castro},
  \bibinfo{person}{Aakash Patel}, \bibinfo{person}{Tom Torfs}, {and}
  \bibinfo{person}{Alper Bozkurt}.} \bibinfo{year}{2021}\natexlab{}.
\newblock \showarticletitle{Noncontact electrophysiology monitoring systems for
  assessment of canine-human interactions}. In \bibinfo{booktitle}{\emph{IEEE
  Sensors}}. \bibinfo{pages}{1--4}.
\newblock


\bibitem[Al-Naji et~al\mbox{.}(2019)]%
        {al2019pilot}
\bibfield{author}{\bibinfo{person}{Ali Al-Naji}, \bibinfo{person}{Yiting Tao},
  \bibinfo{person}{Ian Smith}, {and} \bibinfo{person}{Javaan Chahl}.}
  \bibinfo{year}{2019}\natexlab{}.
\newblock \showarticletitle{A pilot study for estimating the cardiopulmonary
  signals of diverse exotic animals using a digital camera}.
\newblock \bibinfo{journal}{\emph{Sensors}} \bibinfo{volume}{19},
  \bibinfo{number}{24} (\bibinfo{year}{2019}), \bibinfo{pages}{5445}.
\newblock


\bibitem[Apple.com(2022)]%
        {iphone6sspecs}
\bibfield{author}{\bibinfo{person}{Apple.com}.}
  \bibinfo{year}{2022}\natexlab{}.
\newblock \bibinfo{title}{iPhone 6s - Technical Specifications}.
\newblock
\newblock
\urldef\tempurl%
\url{https://support.apple.com/kb/SP726}
\showURL{%
Retrieved May 1, 2022 from \tempurl}


\bibitem[Barbosa~Pereira et~al\mbox{.}(2019)]%
        {barbosa2019contactless}
\bibfield{author}{\bibinfo{person}{Carina Barbosa~Pereira},
  \bibinfo{person}{Henriette Dohmeier}, \bibinfo{person}{Janosch Kunczik},
  \bibinfo{person}{Nadine Hochhausen}, \bibinfo{person}{Ren{\'e} Tolba}, {and}
  \bibinfo{person}{Michael Czaplik}.} \bibinfo{year}{2019}\natexlab{}.
\newblock \showarticletitle{Contactless monitoring of heart and respiratory
  rate in anesthetized pigs using infrared thermography}.
\newblock \bibinfo{journal}{\emph{{PLOS ONE}}} \bibinfo{volume}{14},
  \bibinfo{number}{11} (\bibinfo{year}{2019}), \bibinfo{pages}{e0224747}.
\newblock


\bibitem[Becker(2006)]%
        {becker2006fundamentals}
\bibfield{author}{\bibinfo{person}{Daniel~E Becker}.}
  \bibinfo{year}{2006}\natexlab{}.
\newblock \showarticletitle{Fundamentals of electrocardiography
  interpretation}.
\newblock \bibinfo{journal}{\emph{Anesthesia Progress}} \bibinfo{volume}{53},
  \bibinfo{number}{2} (\bibinfo{year}{2006}), \bibinfo{pages}{53--64}.
\newblock


\bibitem[Chen and McDuff(2018)]%
        {chen2018deepphys}
\bibfield{author}{\bibinfo{person}{Weixuan Chen} {and} \bibinfo{person}{Daniel
  McDuff}.} \bibinfo{year}{2018}\natexlab{}.
\newblock \showarticletitle{{DeepPhys}: Video-based physiological measurement
  using convolutional attention networks}. In
  \bibinfo{booktitle}{\emph{European Conference on Computer Vision}}.
  \bibinfo{pages}{349--365}.
\newblock


\bibitem[Csoltova et~al\mbox{.}(2017)]%
        {csoltova2017behavioral}
\bibfield{author}{\bibinfo{person}{Erika Csoltova},
  \bibinfo{person}{Micha{\"e}l Martineau}, \bibinfo{person}{Alain Boissy},
  {and} \bibinfo{person}{Caroline Gilbert}.} \bibinfo{year}{2017}\natexlab{}.
\newblock \showarticletitle{Behavioral and physiological reactions in dogs to a
  veterinary examination: Owner-dog interactions improve canine well-being}.
\newblock \bibinfo{journal}{\emph{Physiology \& Behavior}}
  \bibinfo{volume}{177} (\bibinfo{year}{2017}), \bibinfo{pages}{270--281}.
\newblock


\bibitem[Csoltova and Mehinagic(2020)]%
        {csoltova2020we}
\bibfield{author}{\bibinfo{person}{Erika Csoltova} {and} \bibinfo{person}{Emira
  Mehinagic}.} \bibinfo{year}{2020}\natexlab{}.
\newblock \showarticletitle{Where do we stand in the domestic dog (Canis
  familiaris) positive-emotion assessment: A state-of-the-art review and future
  directions}.
\newblock \bibinfo{journal}{\emph{Frontiers in Psychology}}
  (\bibinfo{year}{2020}), \bibinfo{pages}{2131}.
\newblock


\bibitem[Dieffenderfer et~al\mbox{.}(2016)]%
        {dieffenderfer2016low}
\bibfield{author}{\bibinfo{person}{James Dieffenderfer}, \bibinfo{person}{Henry
  Goodell}, \bibinfo{person}{Steven Mills}, \bibinfo{person}{Michael McKnight},
  \bibinfo{person}{Shanshan Yao}, \bibinfo{person}{Feiyan Lin},
  \bibinfo{person}{Eric Beppler}, \bibinfo{person}{Brinnae Bent},
  \bibinfo{person}{Bongmook Lee}, \bibinfo{person}{Veena Misra},
  {et~al\mbox{.}}} \bibinfo{year}{2016}\natexlab{}.
\newblock \showarticletitle{Low-power wearable systems for continuous
  monitoring of environment and health for chronic respiratory disease}.
\newblock \bibinfo{journal}{\emph{IEEE Journal of Biomedical and Health
  Informatics}} \bibinfo{volume}{20}, \bibinfo{number}{5}
  (\bibinfo{year}{2016}), \bibinfo{pages}{1251--1264}.
\newblock


\bibitem[Foster et~al\mbox{.}(2018a)]%
        {foster2018system}
\bibfield{author}{\bibinfo{person}{Marc Foster}, \bibinfo{person}{Eric
  Beppler}, \bibinfo{person}{Timothy Holder}, \bibinfo{person}{James
  Dieffenderfer}, \bibinfo{person}{Patrick Erb}, \bibinfo{person}{Kristy
  Everette}, \bibinfo{person}{Margaret Gruen}, \bibinfo{person}{Tamara Somers},
  \bibinfo{person}{Tom Evans}, \bibinfo{person}{Michael Daniele},
  \bibinfo{person}{David~L. Roberts}, {and} \bibinfo{person}{Alper Bozkurt}.}
  \bibinfo{year}{2018}\natexlab{a}.
\newblock \showarticletitle{A system for assessment of canine-human interaction
  during animal-assisted therapies}. In \bibinfo{booktitle}{\emph{40th Annual
  International Conference of the IEEE Engineering in Medicine and Biology
  Society (EMBC)}}. \bibinfo{pages}{4347--4350}.
\newblock


\bibitem[Foster et~al\mbox{.}(2018b)]%
        {foster20183d}
\bibfield{author}{\bibinfo{person}{Marc Foster}, \bibinfo{person}{Patrick Erb},
  \bibinfo{person}{Brenna Plank}, \bibinfo{person}{Helen West},
  \bibinfo{person}{Jane Russenberger}, \bibinfo{person}{Margaret Gruen},
  \bibinfo{person}{Michael Daniele}, \bibinfo{person}{David~L Roberts}, {and}
  \bibinfo{person}{Alper Bozkurt}.} \bibinfo{year}{2018}\natexlab{b}.
\newblock \showarticletitle{3D-printed electrocardiogram electrodes for heart
  rate detection in canines}. In \bibinfo{booktitle}{\emph{IEEE Biomedical
  Circuits and Systems Conference (BioCAS)}}. \bibinfo{pages}{1--4}.
\newblock


\bibitem[Foster et~al\mbox{.}(2019)]%
        {foster2019preliminary}
\bibfield{author}{\bibinfo{person}{Marc Foster}, \bibinfo{person}{Sean Mealin},
  \bibinfo{person}{Margaret Gruen}, \bibinfo{person}{David~L Roberts}, {and}
  \bibinfo{person}{Alper Bozkurt}.} \bibinfo{year}{2019}\natexlab{}.
\newblock \showarticletitle{Preliminary evaluation of a wearable sensor system
  for assessment of heart rate, heart rate variability, and activity level in
  working dogs}. In \bibinfo{booktitle}{\emph{IEEE Sensors}}.
  \bibinfo{pages}{1--4}.
\newblock


\bibitem[Giordano(2022)]%
        {best2022veterinaryservices}
\bibfield{author}{\bibinfo{person}{Medea Giordano}.}
  \bibinfo{year}{2022}\natexlab{}.
\newblock \bibinfo{title}{The Best Veterinary Telemedicine Services for Your
  Pet (2022) | WIRED}.
\newblock
\newblock
\urldef\tempurl%
\url{https://www.wired.com/story/best-veterinary-telemedicine-services/}
\showURL{%
Retrieved May 1, 2022 from \tempurl}


\bibitem[Holder et~al\mbox{.}(2020a)]%
        {holder2020systematicpart1}
\bibfield{author}{\bibinfo{person}{Timothy~RN Holder},
  \bibinfo{person}{Margaret~E Gruen}, \bibinfo{person}{David~L Roberts},
  \bibinfo{person}{Tamara Somers}, {and} \bibinfo{person}{Alper Bozkurt}.}
  \bibinfo{year}{2020}\natexlab{a}.
\newblock \showarticletitle{A systematic literature review of animal-assisted
  interventions in oncology (Part I): Methods and results}.
\newblock \bibinfo{journal}{\emph{Integrative Cancer Therapies}}
  \bibinfo{volume}{19} (\bibinfo{year}{2020}).
\newblock


\bibitem[Holder et~al\mbox{.}(2020b)]%
        {holder2020systematicpart2}
\bibfield{author}{\bibinfo{person}{Timothy~RN Holder},
  \bibinfo{person}{Margaret~E Gruen}, \bibinfo{person}{David~L Roberts},
  \bibinfo{person}{Tamara Somers}, {and} \bibinfo{person}{Alper Bozkurt}.}
  \bibinfo{year}{2020}\natexlab{b}.
\newblock \showarticletitle{A systematic literature review of animal-assisted
  interventions in oncology (Part II): Theoretical mechanisms and frameworks}.
\newblock \bibinfo{journal}{\emph{Integrative Cancer Therapies}}
  \bibinfo{volume}{19} (\bibinfo{year}{2020}).
\newblock


\bibitem[Jiang et~al\mbox{.}(2021)]%
        {jiang2021learning}
\bibfield{author}{\bibinfo{person}{Shihao Jiang}, \bibinfo{person}{Dylan
  Campbell}, \bibinfo{person}{Yao Lu}, \bibinfo{person}{Hongdong Li}, {and}
  \bibinfo{person}{Richard Hartley}.} \bibinfo{year}{2021}\natexlab{}.
\newblock \showarticletitle{Learning to estimate hidden motions with global
  motion aggregation}. In \bibinfo{booktitle}{\emph{IEEE/CVF International
  Conference on Computer Vision}}. \bibinfo{pages}{9772--9781}.
\newblock


\bibitem[Jorquera-Chavez et~al\mbox{.}(2019)]%
        {jorquera2019modelling}
\bibfield{author}{\bibinfo{person}{Maria Jorquera-Chavez},
  \bibinfo{person}{Sigfredo Fuentes}, \bibinfo{person}{Frank~R Dunshea},
  \bibinfo{person}{Robyn~D Warner}, \bibinfo{person}{Tomas Poblete}, {and}
  \bibinfo{person}{Ellen~C Jongman}.} \bibinfo{year}{2019}\natexlab{}.
\newblock \showarticletitle{Modelling and validation of computer vision
  techniques to assess heart rate, eye temperature, ear-base temperature and
  respiration rate in cattle}.
\newblock \bibinfo{journal}{\emph{Animals}} \bibinfo{volume}{9},
  \bibinfo{number}{12} (\bibinfo{year}{2019}), \bibinfo{pages}{1089}.
\newblock


\bibitem[Jukan et~al\mbox{.}(2017)]%
        {jukan2017smart}
\bibfield{author}{\bibinfo{person}{Admela Jukan}, \bibinfo{person}{Xavi
  Masip-Bruin}, {and} \bibinfo{person}{Nina Amla}.}
  \bibinfo{year}{2017}\natexlab{}.
\newblock \showarticletitle{Smart computing and sensing technologies for animal
  welfare: A systematic review}.
\newblock \bibinfo{journal}{\emph{ACM Computing Surveys (CSUR)}}
  \bibinfo{volume}{50}, \bibinfo{number}{1} (\bibinfo{year}{2017}),
  \bibinfo{pages}{1--27}.
\newblock


\bibitem[Lee et~al\mbox{.}(2006)]%
        {lee2006mobile}
\bibfield{author}{\bibinfo{person}{Shang~Ping Lee},
  \bibinfo{person}{Adrian~David Cheok}, \bibinfo{person}{Teh Keng~Soon James},
  \bibinfo{person}{Goh Pae~Lyn Debra}, \bibinfo{person}{Chio~Wen Jie},
  \bibinfo{person}{Wang Chuang}, {and} \bibinfo{person}{Farzam Farbiz}.}
  \bibinfo{year}{2006}\natexlab{}.
\newblock \showarticletitle{A mobile pet wearable computer and mixed reality
  system for human--poultry interaction through the internet}.
\newblock \bibinfo{journal}{\emph{Personal and Ubiquitous Computing}}
  \bibinfo{volume}{10}, \bibinfo{number}{5} (\bibinfo{year}{2006}),
  \bibinfo{pages}{301--317}.
\newblock


\bibitem[Lenovo.com(2022a)]%
        {lenevoflex3specs}
\bibfield{author}{\bibinfo{person}{Lenovo.com}.}
  \bibinfo{year}{2022}\natexlab{a}.
\newblock \bibinfo{title}{Lenovo Flex 3 (1120) Platform Specifications}.
\newblock
\newblock
\urldef\tempurl%
\url{https://psref.lenovo.com/syspool/Sys/PDF/IdeaPad/Lenovo_Flex_3_11/Lenovo_Flex_3_11_Spec.pdf}
\showURL{%
Retrieved May 1, 2022 from \tempurl}


\bibitem[Lenovo.com(2022b)]%
        {lenevoflex3}
\bibfield{author}{\bibinfo{person}{Lenovo.com}.}
  \bibinfo{year}{2022}\natexlab{b}.
\newblock \bibinfo{title}{Product Overview - Lenovo Flex 3 - Lenovo Support
  US}.
\newblock
\newblock
\urldef\tempurl%
\url{https://pcsupport.lenovo.com/us/en/products/laptops-and-netbooks/flex-series/flex-3-1470/solutions/pd100793-product-overview-lenovo-flex-3}
\showURL{%
Retrieved May 1, 2022 from \tempurl}


\bibitem[Li et~al\mbox{.}(2014)]%
        {li2014remote}
\bibfield{author}{\bibinfo{person}{Xiaobai Li}, \bibinfo{person}{Jie Chen},
  \bibinfo{person}{Guoying Zhao}, {and} \bibinfo{person}{Matti Pietikainen}.}
  \bibinfo{year}{2014}\natexlab{}.
\newblock \showarticletitle{Remote heart rate measurement from face videos
  under realistic situations}. In \bibinfo{booktitle}{\emph{IEEE Conference on
  Computer Vision and Pattern Recognition}}. \bibinfo{pages}{4264--4271}.
\newblock


\bibitem[Liu(2009)]%
        {liu2009beyond}
\bibfield{author}{\bibinfo{person}{Ce Liu}.} \bibinfo{year}{2009}\natexlab{}.
\newblock \emph{\bibinfo{title}{Beyond pixels: Exploring new representations
  and applications for motion analysis}}.
\newblock \bibinfo{thesistype}{Ph.\,D. Dissertation}.
  \bibinfo{school}{Massachusetts Institute of Technology}.
\newblock


\bibitem[Maria et~al\mbox{.}(2019)]%
        {maria2019computer}
\bibfield{author}{\bibinfo{person}{Jorquera-Chavez Maria},
  \bibinfo{person}{Fuentes Sigfredo}, {et~al\mbox{.}}}
  \bibinfo{year}{2019}\natexlab{}.
\newblock \showarticletitle{Computer vision and remote sensing to assess
  physiological responses of cattle to pre-slaughter stress, and its impact on
  beef quality: A review}.
\newblock \bibinfo{journal}{\emph{Meat Science}}  \bibinfo{volume}{156}
  (\bibinfo{year}{2019}), \bibinfo{pages}{11--22}.
\newblock
\showISSN{0309-1740}


\bibitem[Mathew et~al\mbox{.}(2021)]%
        {mathew2021remote}
\bibfield{author}{\bibinfo{person}{Joshua Mathew}, \bibinfo{person}{Xin Tian},
  \bibinfo{person}{Min Wu}, {and} \bibinfo{person}{Chau-Wai Wong}.}
  \bibinfo{year}{2021}\natexlab{}.
\newblock \bibinfo{title}{Remote blood oxygen estimation from videos using
  neural networks}.
\newblock
\newblock
\showeprint[arxiv]{2107.05087}


\bibitem[O'Haire(2010)]%
        {o2010companion}
\bibfield{author}{\bibinfo{person}{Marguerite O'Haire}.}
  \bibinfo{year}{2010}\natexlab{}.
\newblock \showarticletitle{Companion animals and human health: Benefits,
  challenges, and the road ahead}.
\newblock \bibinfo{journal}{\emph{Journal of Veterinary Behavior}}
  \bibinfo{volume}{5}, \bibinfo{number}{5} (\bibinfo{year}{2010}),
  \bibinfo{pages}{226--234}.
\newblock


\bibitem[Pereira et~al\mbox{.}(2019)]%
        {pereira2019perspective}
\bibfield{author}{\bibinfo{person}{Carina~B Pereira}, \bibinfo{person}{Janosch
  Kunczik}, \bibinfo{person}{Andr{\'e} Bleich}, \bibinfo{person}{Christine
  Haeger}, \bibinfo{person}{Fabian Kiessling}, \bibinfo{person}{Thomas Thum},
  \bibinfo{person}{Ren{\'e} Tolba}, \bibinfo{person}{Ute Lindauer},
  \bibinfo{person}{Stefan Treue}, {and} \bibinfo{person}{Michael Czaplik}.}
  \bibinfo{year}{2019}\natexlab{}.
\newblock \showarticletitle{Perspective review of optical imaging in welfare
  assessment in animal-based research}.
\newblock \bibinfo{journal}{\emph{Journal of Biomedical Optics}}
  \bibinfo{volume}{24}, \bibinfo{number}{7} (\bibinfo{year}{2019}),
  \bibinfo{pages}{070601}.
\newblock


\bibitem[Ricardo Nathaniel~Holder et~al\mbox{.}(2021)]%
        {ricardo2021ideation}
\bibfield{author}{\bibinfo{person}{Timothy Ricardo Nathaniel~Holder},
  \bibinfo{person}{Evan Williams}, \bibinfo{person}{Devon Martin},
  \bibinfo{person}{Alice Kligerman}, \bibinfo{person}{Emily Summers},
  \bibinfo{person}{Zach Cleghern}, \bibinfo{person}{James Dieffenderfer},
  \bibinfo{person}{Jane Russenberger}, \bibinfo{person}{David Roberts}, {and}
  \bibinfo{person}{Alper Bozkurt}.} \bibinfo{year}{2021}\natexlab{}.
\newblock \showarticletitle{From ideation to deployment: A narrative case study
  of citizen science supported wearables for raising guide dogs}. In
  \bibinfo{booktitle}{\emph{Eight International Conference on Animal-Computer
  Interaction}}. \bibinfo{pages}{1--13}.
\newblock


\bibitem[Rizzo et~al\mbox{.}(2017)]%
        {rizzo2017monitoring}
\bibfield{author}{\bibinfo{person}{Maria Rizzo}, \bibinfo{person}{Francesca
  Arfuso}, \bibinfo{person}{Daniela Alberghina}, \bibinfo{person}{Elisabetta
  Giudice}, \bibinfo{person}{Matteo Gianesella}, {and}
  \bibinfo{person}{Giuseppe Piccione}.} \bibinfo{year}{2017}\natexlab{}.
\newblock \showarticletitle{Monitoring changes in body surface temperature
  associated with treadmill exercise in dogs by use of infrared methodology}.
\newblock \bibinfo{journal}{\emph{Journal of Thermal Biology}}
  \bibinfo{volume}{69} (\bibinfo{year}{2017}), \bibinfo{pages}{64--68}.
\newblock


\bibitem[Samsung.com(2022)]%
        {galaxys10specs}
\bibfield{author}{\bibinfo{person}{Samsung.com}.}
  \bibinfo{year}{2022}\natexlab{}.
\newblock \bibinfo{title}{Camera specifications on the Galaxy S10 | Samsung
  Levant}.
\newblock
\newblock
\urldef\tempurl%
\url{https://www.samsung.com/levant/support/mobile-devices/camera-specifications-on-the-galaxy-s10/}
\showURL{%
Retrieved May 1, 2022 from \tempurl}


\bibitem[Schuurmans et~al\mbox{.}(2020)]%
        {schuurmans2020validity}
\bibfield{author}{\bibinfo{person}{Angela~AT Schuurmans},
  \bibinfo{person}{Peter de Looff}, \bibinfo{person}{Karin~S Nijhof},
  \bibinfo{person}{Catarina Rosada}, \bibinfo{person}{Ron~HJ Scholte},
  \bibinfo{person}{Arne Popma}, {and} \bibinfo{person}{Roy Otten}.}
  \bibinfo{year}{2020}\natexlab{}.
\newblock \showarticletitle{Validity of the Empatica E4 wristband to measure
  heart rate variability (HRV) parameters: A comparison to electrocardiography
  (ECG)}.
\newblock \bibinfo{journal}{\emph{Journal of Medical Systems}}
  \bibinfo{volume}{44}, \bibinfo{number}{11} (\bibinfo{year}{2020}),
  \bibinfo{pages}{1--11}.
\newblock


\bibitem[Singh et~al\mbox{.}(2018)]%
        {singh2018heart}
\bibfield{author}{\bibinfo{person}{Nikhil Singh}, \bibinfo{person}{Kegan~James
  Moneghetti}, \bibinfo{person}{Jeffrey~Wilcox Christle},
  \bibinfo{person}{David Hadley}, \bibinfo{person}{Victor Froelicher}, {and}
  \bibinfo{person}{Daniel Plews}.} \bibinfo{year}{2018}\natexlab{}.
\newblock \showarticletitle{Heart rate variability: An old metric with new
  meaning in the era of using mhealth technologies for health and exercise
  training guidance. Part two: Prognosis and training}.
\newblock \bibinfo{journal}{\emph{Arrhythmia \& Electrophysiology Review}}
  \bibinfo{volume}{7}, \bibinfo{number}{4} (\bibinfo{year}{2018}),
  \bibinfo{pages}{247}.
\newblock


\bibitem[Teed and Deng(2020)]%
        {teed2020raft}
\bibfield{author}{\bibinfo{person}{Zachary Teed} {and} \bibinfo{person}{Jia
  Deng}.} \bibinfo{year}{2020}\natexlab{}.
\newblock \showarticletitle{RAFT: Recurrent all-pairs field transforms for
  optical flow}. In \bibinfo{booktitle}{\emph{European Conference on Computer
  Vision}}. Springer, \bibinfo{pages}{402--419}.
\newblock


\bibitem[Tian et~al\mbox{.}(2022)]%
        {tian2022multi}
\bibfield{author}{\bibinfo{person}{Xin Tian}, \bibinfo{person}{Chau-Wai Wong},
  \bibinfo{person}{Sushant~M Ranadive}, {and} \bibinfo{person}{Min Wu}.}
  \bibinfo{year}{2022}\natexlab{}.
\newblock \showarticletitle{A multi-channel ratio-of-ratios method for
  noncontact hand video based SpO$_2$ monitoring using smartphone cameras}.
\newblock \bibinfo{journal}{\emph{IEEE Journal of Selected Topics in Signal
  Processing}} \bibinfo{volume}{16}, \bibinfo{number}{2}
  (\bibinfo{year}{2022}), \bibinfo{pages}{197--207}.
\newblock


\bibitem[Wang et~al\mbox{.}(2021)]%
        {wang2021contactless}
\bibfield{author}{\bibinfo{person}{Meiqing Wang}, \bibinfo{person}{Ali
  Youssef}, \bibinfo{person}{Mona Larsen}, \bibinfo{person}{Jean-Loup Rault},
  \bibinfo{person}{Daniel Berckmans}, \bibinfo{person}{Jeremy~N
  Marchant-Forde}, \bibinfo{person}{Joerg Hartung}, \bibinfo{person}{Andr{\'e}
  Bleich}, \bibinfo{person}{Mingzhou Lu}, {and} \bibinfo{person}{Tomas
  Norton}.} \bibinfo{year}{2021}\natexlab{}.
\newblock \showarticletitle{Contactless video-based heart rate monitoring of a
  resting and an anesthetized pig}.
\newblock \bibinfo{journal}{\emph{Animals}} \bibinfo{volume}{11},
  \bibinfo{number}{2} (\bibinfo{year}{2021}), \bibinfo{pages}{442}.
\newblock


\bibitem[Wang et~al\mbox{.}(2019a)]%
        {wang2019distinction}
\bibfield{author}{\bibinfo{person}{Pengfei Wang}, \bibinfo{person}{Fulai
  Liang}, \bibinfo{person}{Yangyang Ma}, \bibinfo{person}{Yang Zhang},
  \bibinfo{person}{Huijun Xue}, \bibinfo{person}{Zhao Li}, \bibinfo{person}{Hao
  Lv}, {and} \bibinfo{person}{Jianqi Wang}.} \bibinfo{year}{2019}\natexlab{a}.
\newblock \showarticletitle{Distinction between human and animal in respiratory
  monitoring based on IR-UWB radar}. In \bibinfo{booktitle}{\emph{IEEE
  Photonics \& Electromagnetics Research Symposium-Fall (PIERS-Fall)}}.
  \bibinfo{pages}{392--396}.
\newblock


\bibitem[Wang et~al\mbox{.}(2020)]%
        {wang2020non}
\bibfield{author}{\bibinfo{person}{Pengfei Wang}, \bibinfo{person}{Yangyang
  Ma}, \bibinfo{person}{Fulai Liang}, \bibinfo{person}{Yang Zhang},
  \bibinfo{person}{Xiao Yu}, \bibinfo{person}{Zhao Li}, \bibinfo{person}{Qiang
  An}, \bibinfo{person}{Hao Lv}, {and} \bibinfo{person}{Jianqi Wang}.}
  \bibinfo{year}{2020}\natexlab{}.
\newblock \showarticletitle{Non-contact vital signs monitoring of dog and cat
  using a UWB radar}.
\newblock \bibinfo{journal}{\emph{Animals}} \bibinfo{volume}{10},
  \bibinfo{number}{2} (\bibinfo{year}{2020}), \bibinfo{pages}{205}.
\newblock


\bibitem[Wang et~al\mbox{.}(2019b)]%
        {wang2019method}
\bibfield{author}{\bibinfo{person}{Pengfei Wang}, \bibinfo{person}{Yang Zhang},
  \bibinfo{person}{Yangyang Ma}, \bibinfo{person}{Fulai Liang},
  \bibinfo{person}{Qiang An}, \bibinfo{person}{Huijun Xue},
  \bibinfo{person}{Xiao Yu}, \bibinfo{person}{Hao Lv}, {and}
  \bibinfo{person}{Jianqi Wang}.} \bibinfo{year}{2019}\natexlab{b}.
\newblock \showarticletitle{Method for distinguishing humans and animals in
  vital signs monitoring using IR-UWB radar}.
\newblock \bibinfo{journal}{\emph{International Journal of Environmental
  Research and Public Health}} \bibinfo{volume}{16}, \bibinfo{number}{22}
  (\bibinfo{year}{2019}), \bibinfo{pages}{4462}.
\newblock


\bibitem[Williams et~al\mbox{.}(2020)]%
        {williams2020smart}
\bibfield{author}{\bibinfo{person}{Evan Williams}, \bibinfo{person}{Zachary
  Cleghern}, \bibinfo{person}{Marc Foster}, \bibinfo{person}{Timothy Holder},
  \bibinfo{person}{David Roberts}, {and} \bibinfo{person}{Alper Bozkurt}.}
  \bibinfo{year}{2020}\natexlab{}.
\newblock \showarticletitle{A smart collar for assessment of activity levels
  and environmental conditions for guide dogs}. In
  \bibinfo{booktitle}{\emph{42nd Annual International Conference of the IEEE
  Engineering in Medicine \& Biology Society (EMBC)}}.
  \bibinfo{pages}{4628--4631}.
\newblock


\bibitem[Zhu et~al\mbox{.}(2018)]%
        {zhu2018adaptiveconf}
\bibfield{author}{\bibinfo{person}{Qiang Zhu}, \bibinfo{person}{Mingliang
  Chen}, \bibinfo{person}{Chau-Wai Wong}, {and} \bibinfo{person}{Min Wu}.}
  \bibinfo{year}{2018}\natexlab{}.
\newblock \showarticletitle{Adaptive multi-trace carving based on dynamic
  programming}. In \bibinfo{booktitle}{\emph{2018 52nd Asilomar Conference on
  Signals, Systems, and Computers}}. IEEE, \bibinfo{pages}{1716--1720}.
\newblock


\bibitem[Zhu et~al\mbox{.}(2020)]%
        {zhu2020adaptive}
\bibfield{author}{\bibinfo{person}{Qiang Zhu}, \bibinfo{person}{Mingliang
  Chen}, \bibinfo{person}{Chau-Wai Wong}, {and} \bibinfo{person}{Min Wu}.}
  \bibinfo{year}{2020}\natexlab{}.
\newblock \showarticletitle{Adaptive multi-trace carving for robust frequency
  tracking in forensic applications}.
\newblock \bibinfo{journal}{\emph{IEEE Transactions on Information Forensics
  and Security}}  \bibinfo{volume}{16} (\bibinfo{year}{2020}),
  \bibinfo{pages}{1174--1189}.
\newblock


\bibitem[Zhu et~al\mbox{.}(2017)]%
        {zhu2017fitness}
\bibfield{author}{\bibinfo{person}{Qiang Zhu}, \bibinfo{person}{Chau-Wai Wong},
  \bibinfo{person}{Chang-Hong Fu}, {and} \bibinfo{person}{Min Wu}.}
  \bibinfo{year}{2017}\natexlab{}.
\newblock \showarticletitle{Fitness heart rate measurement using face videos}.
  In \bibinfo{booktitle}{\emph{IEEE International Conference on Image
  Processing}}. \bibinfo{pages}{2000--2004}.
\newblock


\end{thebibliography}

\end{document}